\documentclass[manuscript=article]{achemso}

\usepackage{amssymb}
\usepackage{amsmath}
\usepackage{color}
\usepackage{subfigure}
\usepackage{graphicx}
\usepackage{appendix}

\usepackage{mciteplus}

\setkeys{acs}{usetitle = true}

\def\@dotsep{4.5}
\makeatother

\usepackage{bm}
\usepackage{graphicx}
\usepackage{amssymb}
\usepackage{amsmath}
\usepackage{hyperref}

\title{ Helium Isotopes Quantum Sieving Through Graphtriyne Membranes}

\author{Marta I.  Hern\'{a}ndez}
\affiliation{Instituto de F\'{\i}sica Fundamental,
Consejo Superior de Investigaciones Cient\'{\i}ficas (IFF-CSIC), Serrano 123,
28006 Madrid, Spain}

\author{Massimiliano Bartolomei}
\affiliation{Instituto de F\'{\i}sica Fundamental,
Consejo Superior de Investigaciones Cient\'{\i}ficas (IFF-CSIC), Serrano 123,
28006 Madrid, Spain}

\author{Jos{\'e} Campos-Mart\'{\i}nez} \email{jcm@iff.csic.es}
\affiliation{Instituto de F\'{\i}sica Fundamental,
Consejo Superior de Investigaciones Cient\'{\i}ficas (IFF-CSIC), Serrano 123,
28006 Madrid, Spain}

\date{\today}

\begin{document}

\maketitle

\begin{abstract}

    We report accurate quantum calculations of the sieving of Helium atoms
  by two-dimensional (2D) graphtriyne layers with a new interaction
  potential.  Thermal rate constants and permeances in an ample temperature
  range are computed and compared for both Helium isotopes.  With a pore
  larger than graphdiyne, the most common member  of the \texorpdfstring{$\gamma-$graphyne}
  family, it could be expected that the appearance
  of quantum effects were more limited.  We find, however, a strong
  quantum behavior that can be attributed to the presence of selective
  adsorption resonances, with a pronounced effect in the low 
  temperature regime. This effect leads to the appearance of 
  some selectivity at very low
  temperatures and the possibility for the heavier isotope to cross the
  membrane more efficiently than the lighter, contrarily to what happened with
  graphdiyne membranes, where the sieving at low energy is predominantly
  ruled by quantum tunneling. The use of
  more approximate methods could be not advisable in these situations
  and prototypical transition state theory treatments might lead
  to large errors. 

\end{abstract}

\maketitle

\newpage

\section{I. Introduction}

Two dimensional (2D) materials have become ubiquitous and are being
increasingly used in many technological applications
\cite{advan-2dmat:17,karin-2dmat:20}
with progress in both theory\cite{Yeo_AM:2019} and
experiment\cite{Jia_ACR:2017}.  The fabrication of
different 2D membranes by the conjunction of several so-called top-down
approaches as well as bottom-up methods are able to produce a layer
with almost any feature at will\cite{karin-2dmat:20}.   From the, nowadays, 
very famous graphene,
a large set of many different compounds are now produced in the form of
monoatomic layer 2D materials.  Among all of these,
graphynes and within them, {$\gamma-$graphynes}
\cite{world-graph2yne:19,graphyne-family:19}   
are very popular since several routes to fabrication have been
developed and many possible technological applications have been 
envisaged\cite{research-graph2yne:18}.
These materials are made solely of carbon atoms bonded by $sp$ and $sp^2$
hybridizations. In simple words they consist of benzene rings joined by
$n-$acetylenic groups, giving rise to increasing nanopore
sizes in graphdiyne for $n=2$, graphtriyne for $n=3$
and so on.  Thus, as an example of the fruitful interplay between
theory and experiment previously mentioned, in this family graphdiyne 
was first proposed theoretically\cite{Baughman_JCP:1987} and synthesized 
later on \cite{Li_CC:2010} for the first time, some years before new procedures 
were designed
\cite{Zhou_JACS:2015,matsu-synt-gr2:17} and the route to lab fabrication
became quite regular and common\cite{Jia_ACR:2017}, with the theoretical
work always side-by-side as an invaluable tool
\cite{Swathi_JPCB:2018,James_RSCA:2018}.

Among the multiple applications\cite{research-graph2yne:18}, 
2D materials were early proposed as efficient sieves at the molecular
level\cite{Nature-nano:2012,Marlies:2013,review-JPCL:2015}, and in
this context the graphyne family was one of the suitable candidates
due to the regularity in the position and size of the nanopores.
Particularly, it has been suggested that these  membranes could be 
efficiently used for the isotopic separation of several atomic or 
molecular species \cite{Schrier:10,Hauser:2012,graph3yne-2lay-sep-mb:18}, a process which continues 
being nowadays very challenging and troublesome.  Helium is a scarce resource, 
mostly obtained from natural gas, from where it needs to be separated from 
the other components.  It is mostly used for applications related to technology
and the lightest isotope, $^3$He, is even scarcer, yet it is critical for 
several technologies, and this demand produces from 
time to time a shortage in the supplies of this species 
\cite{ScienceHe3:09,Nature-He:2012}. 
The need for isotopic separation in other simple gases is also
important, as for example in the case of hydrogen \cite{RSCAdv:2012}, so
the need for efficient and cheap filters at the molecular level  is 
expected to increase in the near feature.

For low enough pressures, the study of the separation of different 
compounds from a gas mixture can be modeled as the dynamics of an atom or 
 molecule interacting with the 2D membrane.  At low temperatures, 
 it is expected that quantum effects may play an important role
 leading to larger selectivity ratios.  
 Recently, a rather large number of works have shown
 that quantum tunneling  might rule an efficient separation process
\cite{Hauser:2012,Schrier:12,ceotto:2014,ourJPCC2014,Hernandez_JPCA:2015,Lalitha2015,Li:2015,Qu2016}.  
While the quantum nature of the processes was stressed, most of the calculations
relied on classical mechanics and/or combinations of one-dimensional
calculations as well as some estimations using Transition State Theory (TST)
\cite{Hernandez_JPCA:2015,Truhlar-jcp:72,Truhlar-jpc:79} based on the calculation
of the zero point energy (ZPE) along the 2D nanopore directions, that
has also been shown to be of importance by theory
\cite{Beenakker:95,hankel-pccp:2011,Bathia:2005,Schrier-cpl2012,Marlies:2013}
and  experiment\cite{Zhao:06,Nguyen2010}.

It is important to emphasize that tunneling and ZPE quantum phenomena
affect different isotopes in opposite directions, thus on the one hand,
the lighter atom or
  molecule will overcome more efficiently a barrier by tunneling than the
  heavier one.   On the other hand, the latter will have a lower ZPE along
  the membrane nanopore and therefore will cross more efficiently to the other
  side of the layer.   We have shown in graphdiyne\cite{Hernandez_JPCA:2015} the
  relation and relative importance of these two effects.  More 
  recently\cite{wp3d-jpcc-17}, the limits of more approximated treatments,
  such as one-dimensional tunneling and TST, were assessed by means of
  three-dimensional wave packet calculations.  We found that one possible
  scenario for these popular
  methodologies to breakdown were those in which there were no barrier
  to impede atoms crossing the membrane, a case that, as we will see
  below, is just happening in graphtriyne for $He$ atoms. 
  
   We here present quantum-mechanical calculations for the transmission
of $He$ atoms through a periodic and rigid one-atom-thick graphtriyne membrane,
with special emphasis in the low energy regime, and the effect that
could be shown in the behavior of its isotopes ($^3He,\; ^4He$).
Graphtriyne is the next member of $\gamma-$graphyne family to the
very much studied graphdiyne, and to the best of our knowledge it has not been
synthesized yet, although the next one by the number of acetylenic groups,
graphtetrayne, has been very recently reported as successfully synthesized
\cite{graph4yne-synthesis:18,grap4yne-synth:20}.  Although it
appears that odd members of the graphyne family could be more difficult
to prepare in the lab, one can not completely rule out the possibility
of a novel procedure that could reach a successful synthesis of this
material.   Therefore, our aim in the study of graphtriyne is twofold.
On the one hand, we want to predict what one can expect of the behavior
and properties of such a material, which presents a nanopore certainly larger
than graphdiyne and, on the other hand, to investigate how the properties change as the nanopore increases
its size.  We anticipate that the behavior is far from simple and that
a strong quantum behavior is present at low energies; which
should be taken into account in future studies whether theoretical
or experimental.  As in previous works\cite{wp3d-jpcc-17} our methodology
combines wavepacket propagation by a periodic surface\cite{Yinnon:83} and the
calculation of transmission probabilities using a flux
surface\cite{Miller:74,Zhang:91} with a new and accurate potential energy
surface computed with a procedure that has already proven to be
reliable\cite{grapheneours:2013} .
 After that, thermal rate coefficients and
the more common, in this context, permeance magnitudes will be
computed and discussed.

The paper is organized as follows. In the following Section we present the
system with the interaction potential and dynamical method to study
the transmission of a 3D wave packet through a periodic
membrane.  Results are presented and discussed in Section III, and finally
in Section IV we end with conclusion and perspectives.


\section{II. Theoretical approach.}

\begin{figure}[H]
\centering
    \subfigure[]{\includegraphics[width=0.49\textwidth]{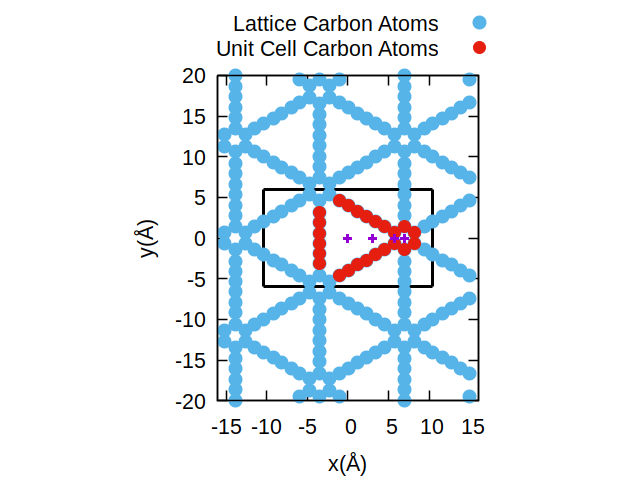}}
    \subfigure[]{\includegraphics[width=0.49\textwidth]{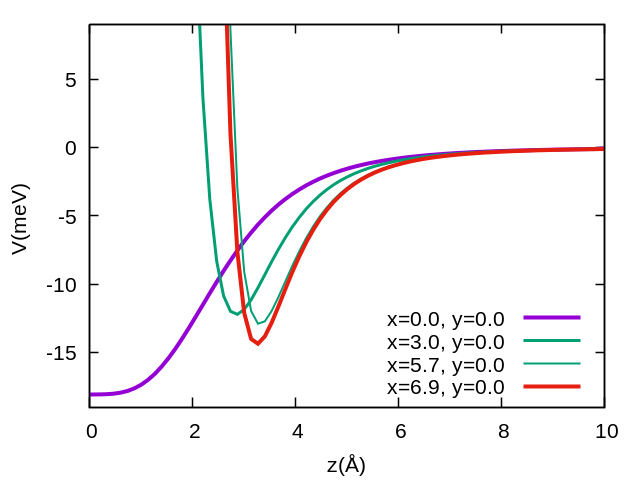}}
\caption{ (a) Graphtriyne lattice.  The red filled points represent the
  atoms of the unit cell and the rectangle indicates the ($x, y$) grid for the
  wave packet propagation.  Crossed points indicate $z$ directions
  at which the interaction potential is computed, as shown in in the right panel.
  (b) Interaction potential along the $z$ coordinate at several $(x,y)$
  positions indicated in the lower right part, and by crossed points in the
  left panel. }
  \label{red-pot}
\end{figure}

We consider the collision of a $He$ atom with a non-vibrating periodic 
single monolayer, i.e. graphtriyne, and we describe the process by solving the
Time-Dependent Sch\"odinger Equation (TDSE), once a suitable interaction 
potential has been computed.  The lattice is represented in
Fig.(\ref{red-pot}-a), where the atoms belonging to the non primitive
(rhomboid) unit cell are
represented by red filled points.  The unit cell is slightly different to
the one commonly used \cite{graph-derivados:14} in that it includes the
middle of the pore. The reason for
this choice is to facilitate the use of a rectangular periodic unit, plotted
as a black rectangle in the same figure, to ease the quantum calculation on
a rectangular grid, as explained below.  When the $He$ is in a given $(x,y,z)$
position it
feels the 2D layer and in Fig.(\ref{red-pot}-b) we show how is this interaction
as we approach the graphtriyne membrane in the direction ($z$) perpendicular
to the surface, in four different positions of the unit cell.  Since the 
nanopore is large enough for the atom size, there is no barrier to surmount
when we are approaching by the middle of the pore.  This interaction potential,
calculated as indicated below, decreases the well and then an expected barrier
appears as we move out of the center of the pore.

\subsection{The interaction potential}

The interaction potential is given by a sum of pairwise interactions between
the $He$ and the carbon atoms present in the layer.   These atom-atom
interactions, depending on $r$, the distance between the rare gas atom and
those in the lattice, are expressed in the so-called  Improved
Lennard-Jones (ILJ)\cite{ILJ}  formula given by

\begin{equation}
V(r)=\varepsilon\left(\frac{6}{n(r)-6}\left(\frac{R_{m}}{r}\right)^{n(r)}-
\frac{n(r)}{n(r)-6}\left(\frac{R_{m}}{r}\right)^{6}\right)
\label{ILJ}
\end{equation}

\noindent
where, $\epsilon$ and $R_m$ are the usual well depth and equilibrium distance
and 

\begin{equation}
n(r)=\beta+4\left(\frac{r}{R_{m}}\right)^{2}
\end{equation}

The parameters of this pair potential were optimized from comparison
with benchmark estimations of interactions energies, obtained at
the ``coupled'' supermolecular second-order
M{\o}ller-Plesset perturbation level of theory\cite{mp2c}, by using
aug-cc-pVTZ and aug-cc-pV5Z basis sets for the carbon and helium atoms,
respectively, and following the guidelines described in detail in
Ref.\cite{ourJPCC2014}.  The values of the parameters are thus given in 
Table \ref{ilj-param}.

\begin{table}[]
\centering
\begin{tabular}{||l||c|c||}
\hline
\hline
      &   He-C              \\
\hline
$R_{m}$ (\AA) & 3.663  \\                       
\hline
$\epsilon$ (meV)   & 1.289          \\                   
\hline
$\beta$ (meV)  &  7.5 \\
\hline
\hline
\end{tabular}
\caption{Parameters of atom-atom ILJ Potential for the interaction of He
  with graphtriyne layer.}
\label{ilj-param}
\end{table}


\subsection{ Wave packet propagation} 

  The (three-dimensional) wave packet is propagated following the
prescriptions given in our early work\cite{wp3d-jpcc-17}, by means of the
Split-Operator method\cite{SplitOperator,K&K:1983}. The membrane spreads 
along $xy$ coordinates, and the atom of mass $\mu$  
is initially represented by a gaussian wave-packet\cite{Heller:75} in
the $z$ direction and 
a plane wave in the remaining ($x,y$) degrees of freedom. In these 
conditions the position of the atom is given by ${\bf r} = ({\bf R}, z)$,
$z$ being the distance to the membrane plane and ${\bf R} = (x,y)$.
The wave packet is discretized on a grid of evenly spaced $(x,y,z)$ points.
To take advantage of the lattice periodicity, the plane wave in the direction 
parallel to the surface is prepared with a wave vector ${\bf K}$ in ${\bf R}$, 
matching the size of the $(x,y)$ grid to the unit cell\cite{Yinnon:83}, 
$(\Delta_x, \Delta_y)$.  
Finally, since the propagation at low energies needs to be carried out for a long period of time, the wave packet is absorbed at the $z-$edge boundaries
\cite{Metiu:87,Pernot:91} by a damping function, to avoid artificial
reflections.  Computational details of the initial wave packet and 
propagation conditions are given in the Appendix.

 The initial wave packet which represents an incident plane wave with 
 a kinetic energy $E=\frac{\hbar^2 k^2}{2\mu}$ and a corresponding wave
 vector ${\bf k}= (k_z,{\bf K})$ is split in a transmitted and a reflected 
wave after reaching the membrane.  The scattering of the wave packet is 
elastic since there is no exchange of energy with the membrane, and 
the parallel wave vectors of these waves obey the Bragg condition whereas
the perpendicular one is modified to satisfy conservation of energy,

\begin{equation}
  k_{z,{\bf G}}^{\pm} = \pm \left[ k^2  - ({\bf K} + {\bf G})^2 \right]^{1/2}
  \label{bragg}
\end{equation}

\noindent
where ${\bf G}$ is a reciprocal lattice vector (see 
Ref.[\cite{wp3d-jpcc-17}] for details).
In these conditions, the probability of transmission,  $P(E)$, through a surface $z= z_{f}$ separating transmitted from incident and reflected waves\cite{Miller:74}, is given by

\vspace*{-0.65cm} 

\begin{equation}
P(E)  = \frac{2 \pi \hbar^2}{\mu} \operatorname{Im} \left( \int dx dy \,
\Psi^{+*}_E(x,y,z_{f}) \frac{d  \Psi^{+}_E}{dz}\mid_{z=z_{f}} \right)
\label{trprob}
\end{equation}

\noindent
where $\Psi^{+*}_E(x,y,z_{f})$ is obtained from the time-energy Fourier 
transform of the evolving wave packet\cite{Zhang:91,cplh2h2:01}. 

The transmission rate coefficient is then obtained from the integration of
$P(E)$, properly weighted by the Boltzmann factor: 

\begin{equation}
R(T) = \frac{1}{h  Q_{trans}} \int e^{-E/(k_BT)} P(E) dE,
\label{eq4}
\end{equation}

\noindent 
where we have changed the more traditional $K(T)$ by $R(T)$ to avoid
confusions with the wave vectors $"k"$ and
is the translational partition function per unit volume, with $k_B$ and $h$ the Boltzmann and Planck constants respectively. In detail,
$P(E)$ would not only depend on the translational energy but also 
of the incident angle, i.e. the parallel wave vector ${\bf K}$. We
are reporting here the results corresponding to a perpendicular
approach ({\bf K}={\bf 0}). 
The effect of varying the angle of incidence ({\bf K}$\neq${\bf 0}) 
would be that of decreasing the value of $P(E)$ so in a sense 
the values reported here should be considered an upper limit.

For the sake of completeness, we have also computed the
permeances\cite{permeance:06}  (effective fluxes per pressure unit
in GPU units ( $1\, GPU=3.35\times10^{-10} \, mol/(m^2 . s .  Pa)$ ),
a magnitude that is more commonly used in sieving and filtering research.
  This magnitude is computed as\cite{Schrier:12,Marlies:2013} ,

\vspace{-0.3cm}

\begin{equation}
  S(T) = \frac{\langle P \rangle_T}{(2 \pi \mu k_B T)^{1/2}} 
\label{eq-permeance}
\end{equation}

\noindent
where $\langle P \rangle_T$ is the thermal average of the probability

\vspace{-0.3cm}

\begin{equation}
  \langle P \rangle_T = \left(\frac{\mu}{2 \pi k_B T}\right) ^{1/2} \int_0^{\infty} P(v_z) \, \exp(\frac{- \mu v_z^2}{2 k_B T}) \, dv_z
\end{equation}

\noindent
and $v_z= ( 2 E / \mu)^{1/2}$ is the velocity of the atom impinging
perpendicularly upon the surface. 

Finally, for both thermal rate coefficients and permeances, selectivity 
is defined as the ratio between the values corresponding to each 
isotope, that is, for isotopes $A$ and $B$, and corresponding 
rates (or permeances)  $M_A$ and $M_B$,  the selectivity 
$S_{A/B} $ is,

\begin{equation}
  S_{A/B} (T) = \frac{M_A(T)}{M_B(T)}.
\label{eq-permeance-selec}
\end{equation}

\begin{figure}[H]
    \centering
\hspace*{-2.cm}\includegraphics[width=8.50cm,angle=0.]{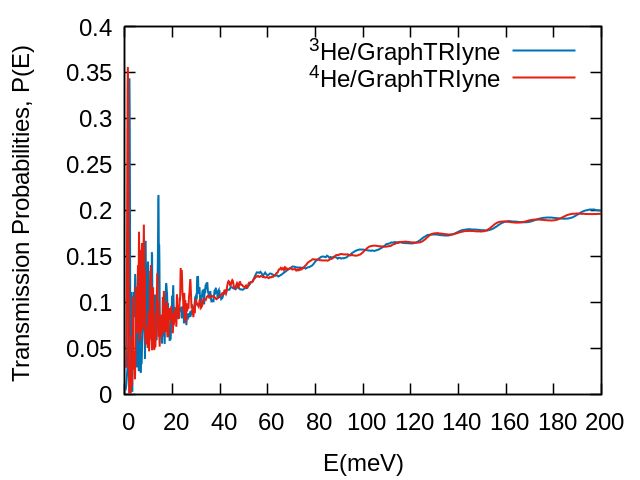}
\caption[] { Transmission probabilities of $^3He$ (blue) and $^4He$ (red)
 as a function of the kinetic energy of  the atoms through graphtriyne. }
  \label{prob-ener}
  \end{figure}

\begin{figure}[H]
    \centering
    \subfigure[]{\includegraphics[width=0.49\textwidth]{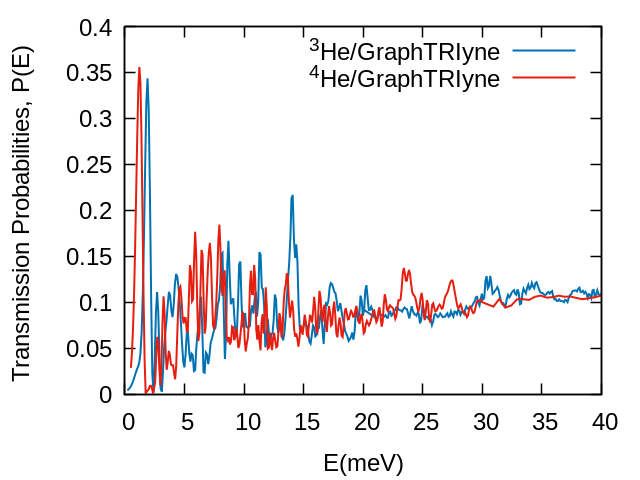}}
\subfigure[]{\includegraphics[width=0.49\textwidth]{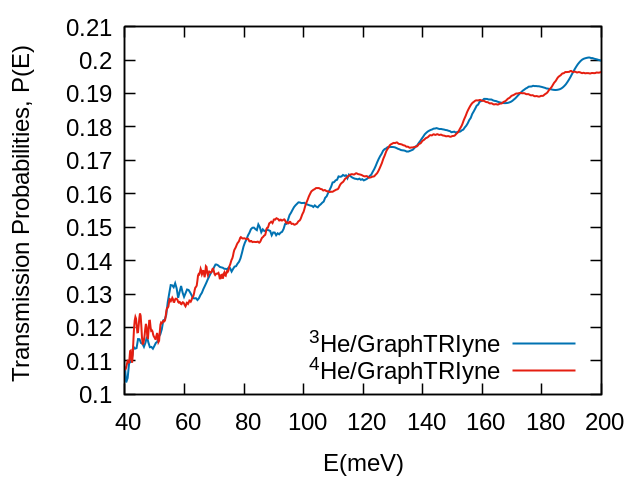}}
\caption{$^4$He and $^3$He transmission probabilities as a function
  of the kinetic energy in (a) low energy regime, (b) at higher
  kinetic energies.}
        \label{prob-ener-low-high}
\end{figure}


\section{III. Results and discussion}

  Once the interaction potential has been presented and following the procedure
described above we have computed the transmission probabilities over a
wide range of initial kinetic energies.  The values of the probabilities are
shown in Fig.(\ref{prob-ener}).  Two different regions can be clearly 
distinguished, one at low energy up to $\approx 40 \, meV$, and thereafter
till the maximum value of incident energy computed.  We have to recall that
for this material the nanopore is large enough that there is no barrier to
impede the passage of $He$ atoms through the middle of it, and that there
is a smooth transition till the acetylenic bonds or benzene ring are
reached, where a barrier begins to take shape and later becomes impenetrable, 
as can be seen in Fig.(\ref{red-pot}-b).
Because of that feature one could expect a smooth behavior already at
low energies, with a transmission probability determined by the effective
pore size at that energy. However by no means this is the case.  
We are presenting in  Fig.(\ref{prob-ener-low-high}-a) a closer look
at the low energy region, where we observe that there is a highly oscillatory
behavior with very strong
peaks at lower energies, in fact larger than values at the maximun
kinetic energies shown in  Fig.(\ref{prob-ener}). This behavior has been checked by changing many
computational features of the wave packet propagation (see Appendix)
to assure that they are not an artifact of the calculations.

 We believe the reason for this oscillatory behavior lies upon another
quantum phenomenon that is not the mentioned tunneling (remember there
is no barrier in the middle of the pore) or the quasi-bound states
in the pore along the $(x,y)$ direction, and that this effect can be appreciated
at higher energies as we will see later on. 
We believe that these numerous peaks in the transmission probability in the
low energy region (Fig.(\ref{prob-ener-low-high}-a)) are due to the so-called
selective adsorption resonances, observed in 1930s gas-surface diffraction
experiments by Stern and coworkers\cite{stern:30} and first explained by
Lennard-Jones and Devonshire\cite{lennard-jones:36} as a temporal trapping
of the atoms in the adsorption well of the gas-surface potential.

These resonances involve a particular case of the Bragg condition of Eq.\ref{bragg},
i.e., when a reciprocal lattice vector ${\mathbf G'}$ produces a transition
to a bound state of the laterally averaged interaction
potential\cite{NuestroPRB:94}. This process can be pictured as a transfer
of momentum from the perpendicular to the parallel direction resulting in
a quasibound state in the perpendicular direction. Due to the corrugation
of the interaction potential, this trapped state is eventually diffracted
at another lattice vector and the particle is scattered back to the
asymptotic region. This feature affects diffraction intensities spectra
by producing  Lorentzian or Fano-type shapes around the energy where the
resonance condition is fulfilled. In the present case of a porous layer,
in addition to being scattered back to the $z>0$ region, the quasibound states
can be leaked through the pores and rather scattered forward towards
the $z<0$ asymptotic region, contributing in this way to an enhancement
of the transmission probability, as observed in
Fig.(\ref{prob-ener-low-high}-a).  These resonances
are different for both isotopes ($^4He, ^3He$), and are located at
different energies.  We note that the first very strong peak corresponds to
the heavier $^4He$ atom and that this will affect macroscopic magnitudes as
we will next see.

In the higher energy range (Fig.(\ref{prob-ener-low-high}-b)) resonant
peaks are  less pronounced and the behavior of the probabilities becomes
somewhat different: they depict a step-like shape, i.e., at certain
energies probabilities rise rapidly and later they smoothly decrease
until the next "step". These jumps in the transmission probabilities
can be understood by the successive population of excited states associated
to the bound motions of the transition state, as
is discussed in detail in Refs. \cite{Hernandez_JPCA:2015,wp3d-jpcc-17}.

The results just presented for transmission probabilities suggest that
because of the strong influence of the resonances, we can expect
some selectiviy in the transmission of one isotope against the other.
{\bf These probabilities are very difficult to converge since it requires
  very long propagation times.  In fact in our case, there is still a small
  portion of the wave packet in the interaction region that could affect
  results at extremely low energy, in the region below $\approx 1 meV$.}

When comparing grahtriyne with other 2D materials with smaller pores (i.e.,
graphdiyne), one would expect a larger flux,
with a smooth increasing behavior as the temperature rises. However,
the presence of a strong resonant behavior at low energies is
very interesting since it 
would allow some selectivity in the quantum sieving, at least at low temperature.
We will see how this manifests in more macroscopic magnitudes.


\begin{figure}[H]
    \centering
    \subfigure[]{\includegraphics[width=0.49\textwidth]{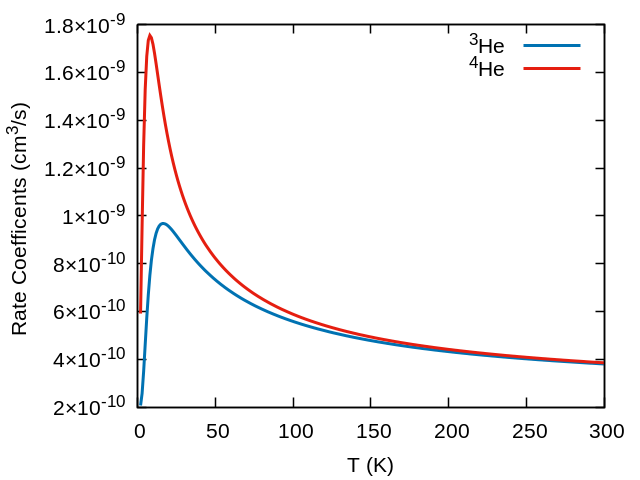}
 }
\subfigure[]{\includegraphics[width=0.49\textwidth]{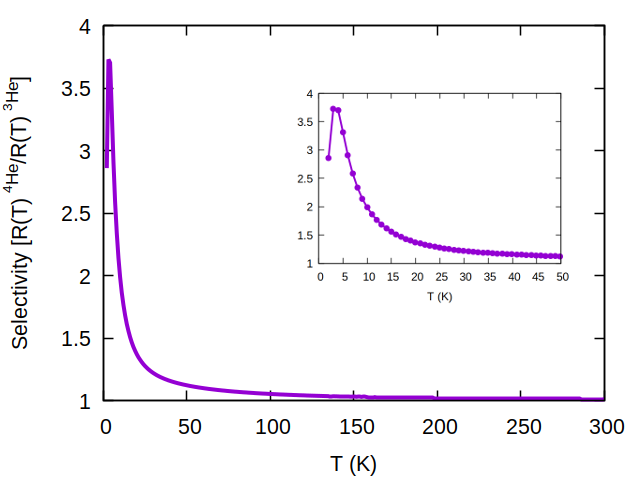}}
\caption{(a) Thermal rate coefficients for the Helium graphtriyne system.
 (b) Rate coefficients  selectivity $^4He/^3He$ .}
        \label{rates}
\end{figure}

 
Rate coefficients as functions of temperature are determined from these
probabilities (Eq.\ref{eq4}), and are shown for both $^3He$ and $^4He$
in Fig.(\ref{rates}-a).  Their values are large as expected
due to the large pore area.  They present a maximum at low temperature
as oppossed to more typical behavior where the rate coefficients increase
with temperture. More noticeable
is the fact that it is the heavier isotope, $^4He$, the one that presents
larger values at low temperature, a different behavior to what
was found for graphdiyne\cite{Hernandez_JPCA:2015,wp3d-jpcc-17} or related
membranes whose behavior at low energy is dominated by the tunneling,
i.e. the ighter isotope, $^3He$, is the one that
it is favored at low temperature.   In Fig.(\ref{rates}-b), we show the ratio
of thermal rates.  As a consequence of the previous effects, we find that 
this system does present a significant selectivity ($\approx 4$), although
for low temperatures as can be
better appreciated in the inset of the figure.

\begin{figure}[h]
    \centering
    \subfigure[]{\includegraphics[width=0.48\textwidth]{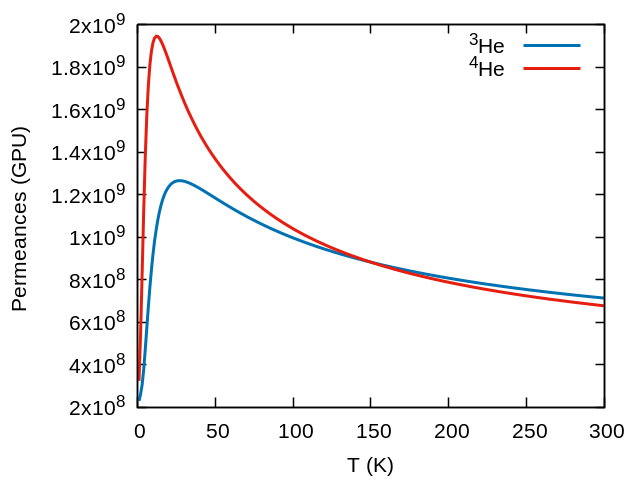}
 }
\subfigure[]{\includegraphics[width=0.48\textwidth]{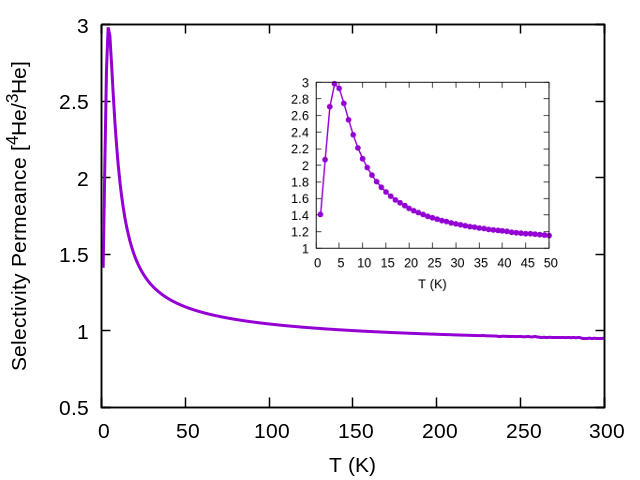} }
\caption{(a) Permeances for $^3He$ and $^4He$ through graphtriyne at
  different temperature. (b)  Permeances selectivity $^4He/^3He$ .}
        \label{permea}
\end{figure}

To finish this section, in Fig.(\ref{permea}-a) we plot the results for the 
computed permeances in GPU units, which as expected, show a similar behavior
to that of rate constants. The corresponding selectivity in Fig.(\ref{permea}-b), also manifests similar values than that
obtained from thermal rates and a maximun yield at about the same temperature.
To finish, we would like to remark 
that it is usually admitted that permeances higher than
$S(T) > 20 \; GPU $ are industrially appealing, a feature that here is clearly
reached for the whole temperature range, up to room temperature.

\section{IV. Conclusions}

We have reported  three-dimensional wave packet calculations for 
the quantum sieving of $He$ atoms through graphtriyne.
The transmission probabilities show a strong oscillatory behavior at
 low energies that leads to isotopic selectivity even though the pore is
large enough to become a barrierless process, where very small or no
selectiviy should be expected.   It is suggested that another
quantum phenomenon, that of selective adsorption resonances, has a strong
influence
over the whole temperature range studied.  This effect, apart from
tunneling or zero point energy effects, has not been considered so far, and
we think it is worth to explore.  Although it will be difficult
to characterize these resonances and to know their features in detail, a
procedure similar to the one
we described in Ref.\cite{NuestroPRB:94}
would help unveil the possibilities of this effect for being used
in this kind of materials and processes.

Rate coefficients and permeances are quite large, as  corresponds with
a material with a large pore size, but exhibiting a maximun and an unexpected selectivity a low temperature.
 The quantum effects at low energies, just commented, are also responsible of a different  behavior to that of graphdiyne regarding the easiness of
transmission, since in this case it is the heavier isotope $^4He$ that
it is favored at low temperature instead the lighter  $^3He$, opening 
new possibilities for the separation of these two isotopes.

Finally, as it has been previously shown\cite{wp3d-jpcc-17}, the absence of
a barrier along the minimun energy path could involve difficulties for approximate treatments based on TST, but
in cases like this, more popular treatments making use of classical mechanics
could also suffer of large inaccuracies because of the pronounced quantum
behavior. It would be then very interesting to
see if the next member of the family, graphtetrayne \cite{graph4yne-synthesis:18,grap4yne-synth:20}, recently synthesized
and with an even larger nanopore, also shows a similar quantum behavior
at low energy and to continue exploring the possibilities of this
quantum effect\cite{NuestroPRB:94,ceotto:2014,h2-cnt-resonan-fermin:17}.

\section*{Acknowledgments}
This work was supported by the Spanish MICINN with Grant
  FIS2017-84391-C2-2-P.
We thank {\it Centro de Supercomputaci\'on de Galicia, CESGA}, for the
allocation of computing time.

\begin{appendices}


\setcounter{table}{0}  
\setcounter{equation}{0}

\appendixpage

\section{Details of wave packet calculations }

The wave packet is discretized on a grid, with values given in
Table(\ref{Grid}), and initially is given as a product of a translational
gaussian wavepacket\cite{Heller:75} for the $z$ coordinate, and a plane wave
for $x,y$ coordinates along the membrane surface

\begin{equation}
  \Psi(\mathbf{r}, t=0)=G\left(z ; z_{0}, k_{z 0},
  \alpha\right) \frac{\exp [i \mathbf{K} \cdot \mathbf{R}]}
              {\left(\Delta_{x} \Delta_{y}\right)^{1 / 2}}
\end{equation}

\noindent
On the one hand, 
$ (\Delta_{x} \Delta_{y} )^{-1 / 2} \exp [i \mathbf{K} \cdot \mathbf{R}]$ is the
wave packet depending on the parallel coordinates,
where $\Delta_{x}$=20.82 \AA \, and $\Delta_{y}$=12.02 \AA \, are the lengths
of the unit cell.  In this work, $\mathbf{K}= \mathbf{0}$
to simulate a wave packet  approaching to the membrane in the perpendicular
direction.  The gaussian wave packet is given by

\begin{equation}
  G(z;z_0,k_{z0},\alpha) = \left( \frac{2
    \operatorname{Im}(\alpha) }{\pi \hbar} \right)^{1/4}
 \exp \left[i \alpha (z-z_0)^2/\hbar + i k_{z0}(z-z_0) \right]  
\label{eq2}
\end{equation}

\noindent
where $z_o$ and $k_{z0}$ are the central values of the wave
packet in the position and momentum spaces, respectively, while $\alpha$ is a
pure imaginary number that determines the width of the gaussian function.

 The grid (and number of points) in the perpendicular
 coordinate is much larger than in the direction of the graphtriyne layer,
 to include regions where the He-membrane interaction is negligible,
 which will be the regions where we will fix the initial wave
 packet and from where absorbing boundary conditions are applied,
 Table(\ref{WP-parameters}).


\begin{table}[H]
\centering
\begin{tabular}{||ccc||c c c|c|c||}
\hline
\hline
     && &   & & $x$     &      $y$  & $z$      \\
\hline
Box (\AA) && & & &  (-10.41,10.41) & (-6.010,6.010) & ( -30.0, 55.0)\\
\hline
Number of points && & & & 128 & 128 & 1024   \\
\hline
\hline
\end{tabular}
\caption{Grid sizes and number of points for the representation of the
  wave packet. Grid boxes for the parallel coordinates correspond to those
  of the unit cell of Fig.(\ref{red-pot}-a)  of the main text.}
\label{Grid}
\end{table}



\small

\begin{table}[H]
\centering
\begin{tabular}{|c|c|c|c|c|c|c|c|c|}
\hline
\multicolumn{3}{|c|}{Gaussian} & \multicolumn{2}{|c|}{Time} & \multicolumn{3}{|c|}{Wave packet} & Flux \\
\multicolumn{3}{|c|}{parameters} & \multicolumn{2}{|c|}{propagation} & \multicolumn{3}{|c|}{splitting} & surface\\
\hline
 $Im(\alpha)\,\frac{[\hbar]}{[L^2]}$ & $z_0$(\AA) & $k_{z_0}$(\AA$^{-1}$) & $\delta t$($fs$) & $t_{final}$($ps$)
 & $\Delta t$($fs$) & $z_+$(\AA) & $\beta$(\AA$^{-2}$) & $z_f$(\AA) \\
\hline
 0.014     & 40.      & [3.3 - 15.0] &   0.083     & [15.0 --18.75] & 0.326 & [45--55] &[0.005-0.05] & [-2.5-- -1.5]  \\
\hline
\end{tabular}
\caption{Parameters of the wave packet propagation, [: - :]
 indicates the range of values taken in different calculations
   (definitions specified in the text).}
\label{WP-parameters}
\end{table}

\normalsize


 Absorbing boundary conditions\cite{Metiu:87,Pernot:91}
were applied for the non-periodic coordinate $(z)$  in the regions
defined by  $z< z_{-}$ and $z> z_+$ and with a time period $\Delta t$, the
wave packet is split into interaction and product wave packets 
using the damping function $\exp{[-\beta(z-z_{\pm})^2]}$. In our particular case,
given the grid size, energy and time propagation, there is no absorpion in
$ z_{-} $, the negative part of the grid along $z$, and the parameters
for the damping function are irrelevant.
After splitting, propagation is resumed using the interaction portion of the
wave packet.

In order to obtain the transmission probability (Eq. 3) it is necessary
to compute the stationary wave function and its derivative at the flux surface $z=z_f$.
This task is performed by accumulating the integrand of the time-energy Fourier
transform along the propagation time\cite{cplh2h2:01}. 
Computations run until most of the wave packet gets out of the
interaction region, remaining only a small portion in this
region ($\approx 1 $\%), for calculations of probabilities shown in
the main text.
Total propagation times $t_{final}$ range between
the values indicated in Table \ref{WP-parameters}, the shortest (largest)
times corresponding to the calculations at largest (smaller) energies
in the initial wavepacket.

Values of all parameters of the calculations mentioned above are gathered in
Table \ref{WP-parameters}. Several convergence checks were carried out by varying
these values as well as the number of grid points and the $z$ box size.

\end{appendices}

\end{document}